\documentclass[10pt]{article}
\usepackage[subpreambles=true, sort=true]{standalone}
\usepackage[a4paper, lmargin=20mm, rmargin=20mm, tmargin=22mm, bmargin=23mm]{geometry}
\usepackage[utf8]{inputenc}
\usepackage[english]{babel}
\usepackage{mlmodern}
\usepackage[font=small]{caption}
\usepackage{bbm}
\usepackage{multicol}
\usepackage[page]{appendix}
\usepackage{titling}
\usepackage{titlesec}
\titleformat{\section}
  {\centering\Large}{\thesection}{1.1em}{}
\titleformat{\subsection}
  {\large}{\thesubsection}{.5em}{}
\titlespacing*{\section}{0pt}{3ex}{3ex}
\titlespacing*{\subsection}{0pt}{3ex}{1ex}

\usepackage{todonotes}
\usepackage{float}
\usepackage{csquotes}
\usepackage{authblk}
\usepackage{amsmath, amsthm, amssymb, amsfonts, mathtools}
\usepackage[backend=biber, citestyle=numeric, sorting=none, maxcitenames=10, url=false, isbn=false]{biblatex}
\usepackage{hyperref}
\hypersetup{
    hidelinks,
    colorlinks=false,
    breaklinks=true,
    pdftitle={Emergent double models}
}
\ifpdf
\else
  \usepackage[hyphenbreaks]{breakurl} 
\fi
\usepackage{eqparbox}
\usepackage{braket}
\providecommand{\dyad}[2]{{\ket{#1}\mkern-6mu\bra{#2}}}
\usepackage{graphicx}
\usepackage[shortlabels]{enumitem}
\setlist[enumerate,1]{label=(\roman*)}
\usepackage{float}
\usepackage[page]{appendix}
\usepackage{tikz}
\usepackage{pgfkeys}
\usepackage{stackengine}
\usepackage{listofitems}
\newcommand\Stk[2]{\genfrac{}{}{0pt}{}{#1}{#2}}
\newcommand\mStk[2][\normalbaselineskip]{%
  \setstackgap{L}{#1}%
  \setsepchar{ }%
  \readlist\inputlist{#2}%
  \raisebox{#1*(\inputlistlen-1)/2}{\lstackMath\Longunderstack[c]{#2}}%
  \setstackgap{L}{\normalbaselineskip}%
}

\newcommand{\eqb}[3][c]{\eqmakebox[#2][#1]{\(\displaystyle #3\)}}
\DeclareMathOperator{\Rep}{\mathsf{Rep}}
\DeclareMathOperator{\Vect}{\mathsf{Vect}}
\DeclareMathOperator{\vect}{\mathsf{vect}}

\newcommand{\tw}[2]{\prescript{#1\mkern-4mu\relax}{}{#2}}

\pgfkeys{
  /TNs/dyad/.cd,
  label/.store in=\TNLabel,
  label left/.store in=\TNLabelLeft,
  label right/.store in=\TNLabelRight,
  width/.store in=\widthTN,
  height/.store in=\heightTN,
  element/.store in=\el,
  pivot/.store in=\RotatePoint,
  element left/.store in=\elLeft,
  element right/.store in=\elRight
}

\newcommand{\tnBra}[1][]{
  \pgfkeys{
    /TNs/dyad/.cd,
    label=\(\), 
    width=1.5cm, 
    element=h,
    #1
  }
  
\begin{tikzpicture}[baseline={([yshift=-\the\fontdimen22\textfont2]D.center)}] 
\node[empty] (D) at (0, 0) {\(\bra{\el}\)};
\node[empty] (Hr) at (\widthTN/3, 0) {\TNLabel};
\draw [mid arrow] (Hr) -- ([xshift=-\the\fontdimen10\textfont3]D.east);
\end{tikzpicture}

}
\newcommand{\tnKet}[1][]{
  \pgfkeys{
    /TNs/dyad/.cd,
    label=\(\), 
    width=1.5cm, 
    element=g,
    #1
  }
  
\begin{tikzpicture}[baseline={([yshift=-\the\fontdimen22\textfont2]D.center)}] 
\node[empty] (D) at (0, 0) {\(\ket{\el}\)};
\node[empty] (Hl) at (-\widthTN/3, 0) {\TNLabel};
\draw [mid arrow] ([xshift=+\the\fontdimen10\textfont3]D.west) -- (Hl);
\end{tikzpicture}

}
\newcommand{\tnKetV}[1][]{
  \pgfkeys{
    /TNs/dyad/.cd,
    label=\(\), 
    width=.5cm, 
    height=1.3cm, 
    pivot=1,
    element=g,
    #1
  }
  
\begin{tikzpicture}[baseline={([yshift=-\the\fontdimen22\textfont2]D.west)}, rotate=-90, transform shape] 
\if1\RotatePoint
  \node[empty] (D) at (0, 0) {\scriptsize\(\Ket{\rotatebox{90}{\(\el\)}}\)};
\else
  \node[empty] (D) at (0, 0) {\scriptsize\(\Ket{\rotatebox[origin = \RotatePoint]{90}{\(\el\)}}\)};
\fi
\node[empty] (Hr) at (-\heightTN/3, 0) {\TNLabel};
\draw[mid arrow] ([xshift=+\the\fontdimen10\textfont3]D.west) -- (Hr);
\node[empty] (lb) at (0, -\widthTN/2) {};
\node[empty] (rb) at (0,  \widthTN/2) {};
\end{tikzpicture}

}
\newcommand{\tnBraV}[1][]{
  \pgfkeys{
    /TNs/dyad/.cd,
    label=\(\), 
    width=.5cm, 
    height=1.3cm, 
    pivot=1,
    element=g,
    #1
  }
  
\begin{tikzpicture}[baseline={([yshift=-\the\fontdimen22\textfont2]D.center)}, rotate=-90, transform shape] 
\if1\RotatePoint
  \node[empty] (D) at (0, 0) {\scriptsize\(\Bra{\rotatebox{90}{\(\el\)}}\)};
\else
  \node[empty] (D) at (0, 0) {\scriptsize\(\Bra{\rotatebox[origin = \RotatePoint]{90}{\(\el\)}}\)};
\fi
\node[empty] (Hr) at (\heightTN/3, 0) {\TNLabel};
\draw[mid arrow] (Hr) -- ([xshift=-\the\fontdimen10\textfont3]D.east);
\node[empty] (lb) at (0, -\widthTN/2) {};
\node[empty] (rb) at (0,  \widthTN/2) {};
\end{tikzpicture}

}
\pgfkeys{
  /TNs/maps/.cd,
  label in/.store in=\TNLabelIn,
  label out/.store in=\TNLabelOut,
  width/.store in=\widthTN,
  color/.store in=\TNColor,
  height/.store in=\heightTN,
  rev/.store in=\revDir,
  rev x/.store in=\revDirX,
  rev y/.store in=\revDirY,
  arrows/.store in=\drawDec,
  corners/.store in=\AngledDom,
  map 1/.store in=\mapA,
  map 2/.store in=\mapB,
  map 3/.store in=\mapC,
  pos 1/.store in=\posA,
  pos 2/.store in=\posB,
  pos 3/.store in=\posC
}

\pgfkeys{
  /TNs/PEPS/.cd,
  width/.store in=\TensorWidth,
  height/.store in=\TensorHeight,
  label/.store in=\TensorLabel,
  label inner/.store in=\TensorLabelInner, 
  pattern/.store in=\TensorPattern,
  map top/.store in=\TopLegOperator,
  map RU/.store in=\RULegOperator,
  map RD/.store in=\RDLegOperator,
  map LU/.store in=\LULegOperator,
  map LD/.store in=\LDLegOperator
}

\pgfkeys{
  /TNs/MPS/.cd,
  width/.store in=\widthMPS,
  color/.store in=\TNColor,
  height/.store in=\heightMPS,
  label/.store in=\MPSTensorLabel,
  label inner/.store in=\MPSTensorLabelInner, 
  label xpos/.store in=\MPSTensorLabelInnerPosX, 
  ulabel/.store in=\MPSTensorULabel,
  ulabel inner/.store in=\MPSTensorULabelInner, 
  index left/.store in=\MPSIndexLeft,
  index right/.store in=\MPSIndexRight,
  index top/.store in=\MPSIndexTop,
  index bottom/.store in=\MPSIndexBottom,
  rep top/.store in=\TopLegOperator,
  rep bottom/.store in=\BottomLegOperator,
  rep left/.store in=\LeftLegOperator,
  rep right/.store in=\RightLegOperator,
  dsep/.store in=\dMPSspace, 
  rep top/.store in=\TopLegOperator,
  map RU/.store in=\RULegOperator,
  map RD/.store in=\RDLegOperator,
  map LU/.store in=\LULegOperator,
  map LD/.store in=\LDLegOperator
}

\theoremstyle{definition}
\newtheorem{theorem}{Theorem}[section]
\newtheorem{theorem-no-proof}[theorem]{Theorem}
\newtheorem{definition}[theorem]{Definition}

\AtEndEnvironment{example}{\null\hfill\ensuremath{\diamond}}
\AtEndEnvironment{theorem-no-proof}{\null\hfill\ensuremath{\square}}

\title{Non-abelian quantum double models from iterated gauging}
\author[1]{David Blanik}
\author[2]{Jos\'e Garre-Rubio}
\affil[1]{University of Vienna, Faculty of Physics, Boltzmanngasse 5, 1090 Vienna, Austria}
\affil[2]{Instituto de F\'isica Te\'orica, UAM-CSIC, C. Nicol\'as Cabrera 13-15, Cantoblanco, 28049 Madrid, Spain}
\date{}
\begin{document}
\renewcommand\Affilfont{\itshape\small}
\renewcommand{\abstractname}{\vspace{-4\baselineskip}}
\maketitle
\begin{abstract}
  \noindent
  We reconstruct all (2+1)D quantum double models of finite groups from their boundary symmetries
  through the repeated application of a gauging procedure,
  extending the existing construction for abelian groups.
  We employ the recently proposed categorical gauging framework,
  based on matrix product operators~(MPOs),
  to derive the appropriate gauging procedure for the \(\Rep G\) symmetries appearing in our construction
  and give an explicit description of the dual emergent \(G\) symmetry,
  which is our main technical contribution.
  Furthermore, we relate the possible gapped boundaries of the quantum double models
  to the quantum phases of the one-dimensional input state to the iterated gauging procedure.
  Finally, we propose a gauging procedure for 1-form \(\Rep G\) symmetries on a two-dimensional lattice and
  use it to extend our results to the construction of (3+1)D quantum doubles models
  through the iterative gauging of (2+1)-dimensional symmetries.
\end{abstract}

\smallskip

\begin{multicols}{2}
\section{Introduction}
\label{sec:introduction}

Dualities and gauging maps are essential to our understanding
of quantum phases of matter in systems with symmetries,
revealing connections between seemingly unrelated models.
For example, the Kramers-Wannier duality~\cite{kramersStatisticsTwoDimensionalFerromagnet1941}
relates the disordered and ordered phases of the quantum Ising model,
precisely identifying its critical point.
Furthermore, these concepts have been instrumental in constructing
and understanding a broad class of exactly solvable lattice models,
most notably Kitaev's quantum double model and its simplest instance,
the toric code \cite{kitaevFaulttolerantQuantumComputation2003a}.

The gauging procedure, introduced by Haegeman et al.~\cite{haegemanGaugingQuantumStates2015},
provides a systematic way to map quantum many-body states,
invariant under a global on-site action of a symmetry group \(G\), to locally symmetric ones.
The procedure introduces auxiliary, so-called \emph{gauge}, degrees of freedom between neighboring sites
and projects onto the subspace satisfying local Gauss law constraints.
Analogous to minimal coupling,
expectation values of symmetric observables are preserved under the gauging procedure.
For a comprehensive discussion of the one-dimensional case in terms of tensor network states
see~\cite{blanikGaugingQuantumPhases2025} and~\cite{williamsonFractalSymmetriesUngauging2016}
for a generalization to subsystem symmetries in higher dimensions.

Crucially, in addition to the local \(G\) symmetry (the Gauss law),
the gauged states exhibit a global categorical \(\Rep G\) symmetry on their gauge degrees of freedom,
dubbed \emph{emergent dual symmetry}.
In the case of abelian \(G\), 
the \(\Rep G\) action can naturally be viewed as a representation of the dual group \(\widehat{G}\)
and the gauging procedure can thus be iterated,
creating a two-dimensional system out of one-dimensional layers of gauge fields.
In~\cite{garre-rubioEmergent2+1DTopological2024} it was demonstrated that in this construction the Gauss laws
map to star and plaquette operators of the (2+1)D quantum double model of \(G\).
This construction has also been extended to higher dimensions and subsystem symmetries,
recovering fracton phases~\cite{cuiperTwistedGaugingTopological2025}.

Extending the construction to non-abelian symmetry groups
requires a suitable generalization of the gauging procedure applicable to global \(\Rep G\) symmetries.
Gauging has been extended to encompass non-invertible symmetries in~\cite{seifnashriGaugingNoninvertibleSymmetries2025}
and a proposal for such a procedure that fits our purposes
was recently made in \cite{cuiperGaugingDualityOnedimensional2025},
extending gauging to arbitrary 1D fusion-categorical symmetries and explaining its connection to dualities.
In their framework, the data defining the gauging procedure, or equivalently the gauged theory,
are encoded as a haploid, symmetric, special Frobenius algebra object internal to the symmetry fusion category:
in the original construction~\cite{haegemanGaugingQuantumStates2015}
corresponds to the group-algebra \(\mathbb{C} G\),
viewed as a Frobenius algebra internal to \(\Vect^G\).

In the present work, we analyze the set of non--Morita~equivalent Frobenius algebras internal to \(\Rep G\),
as classified in \cite{ostrikModuleCategoriesWeak2003, etingofSemisimpleGEquivariantSimple2017},
to establish the appropriate choice of Frobenius algebra
to reproduce the emergence of the quantum double models from the iterated gauging procedure,
also in the case of non-abelian groups.
The crucial step here is to determine explicitly the form of the emergent dual group symmetry,
which in previous works was not clear without disentangling matter from the gauge degrees of freedom.
This allows us to extend the iterative gauging procedure to all on-site symmetries of finite groups,
obtaining the corresponding quantum double models with boundary.
Moreover, we propose a procedure to gauge 1-form \(\Rep G\) symmetries in (2+1)-dimensional systems,
allowing us to extend our construction to obtain the (3+1)D quantum double models. 

We confirm that the proposed gauging procedure for fusion-categorical symmetries
is a practical generalization of the classic gauging procedure and
in our construction we relate the gapped boundaries of the quantum double model to
the quantum phases of the one-dimensional input states to the iterated gauging procedure,
both of which are classified by pairs \((K, \alpha)\),
where \(K \leq G\) and \(\alpha \in H^2(K,U(1))\)
with the additional requirement of \(K\trianglelefteq G\) in the latter case.
Beyond its conceptual value, our approach using gauging could provide a concrete route for simulating and engineering non-abelian topological phases as in~\cite{lyonsProtocolsCreatingAnyons2025}, which are central to proposals for fault-tolerant quantum computation and quantum error correction~\cite{millerResourceQualitySymmetryProtected2015, cuiKitaevsQuantumDouble2020}.

We begin in Section 2 by reviewing the relevant background, namely
the standard gauging procedure for global group symmetries and
the categorical gauging procedure.
In Section 3, we discuss the gauging procedure for \(\Rep G\) symmetries in more detail
and establish the correct choice of Frobenius algebra for use in the iterated gauging procedure,
which we then carry out in Section 4, where we also show the emergence of quantum double models.
Section 5 extends our results to symmetries of (2+1)D systems, yielding (3+1)D quantum double models.
    
\section{Background}

In this section we give a brief summary of the concepts that are relevant to the later sections.
\subsection{Gauging global group symmetries}
\label{sec:gauging-group}
Gauging, as introduced by Haegeman et al.~\cite{haegemanGaugingQuantumStates2015},
is a procedure that transforms a quantum state \(\ket{\psi}\),
invariant under the global, on-site action of a finite group \(G\),
into a locally invariant one, 
while preserving its expectation values with respect to (appropriately dressed) symmetric operators.

Imagine \(\ket{\psi}\) as living on a periodic, one-dimensional lattice, with sites labeled by
\(i \in \mathbb{Z}_n \equiv \mathbb{Z}/n\mathbb{Z}\)
and associated to local spin degrees of freedom, referred to as \emph{matter} degrees of freedom and
described by the Hilbert space \(\mathcal{H}_i\cong \mathbb{C}^{d_i}\),
carrying a unitary representation \(U_i\colon G\rightarrow \operatorname{GL}(\mathcal{H}_i)\)
of the group \(G\).
The state being globally symmetric means that
\(\bigotimes_i U_i(g)\ket{\psi} = \ket{\psi}\) for all \(g \in G\).
While we will set \(U_i = U\) for all \(i\in \mathbb{Z}_n\) going forward, we will continue to include the subscript in cases where it increases clarity.

We begin by introducing new (so-called \emph{gauge}) degrees of freedom,
associated to the edges of the lattice and described by the group algebra \(\mathbb{C}G\).
Crucially, the group algebra carries two commuting representations of \(G\),
\begin{align}
  \label{eq:55}
  L(g) \ket{h}=\ket{gh}, && R(g)\ket{h}=\ket{hg^{-1}},
\end{align}
referred to as the left and the right regular representations of \(G\), respectively,
where we use \(\ket{g}\) for \(g\in G\) to denote the canonical basis vectors of \(\mathbb{C}G\).
The regular representations are used to construct the local symmetry operators
\(\widehat{U}_i(g)\coloneq R(g) \otimes U_i(g) \otimes L(g)\)
acting on the enlarged Hilbert space.
The projection onto the symmetric subspace of \(\widehat{U}_i\) is just
\begin{equation}
  \label{eq:47}
  P_i \coloneq \frac{1}{|G|}\sum_{g\in G} R(g) \otimes U_i(g) \otimes L(g),
\end{equation}
which can be thought of as enforcing a local Gauss law.
Since \(L\) and \(R\) commute,
the local projections \(P_i\) can be combined into the orthogonal projection 
\begin{equation}
  \label{eq:51}
  P \coloneq \prod_{i\in \mathbb{Z}_n} P_i
\end{equation}
onto the subspace invariant under \(\widehat{U}_i\) for all \(i\in \mathbb{Z}_n\).
Finally, to produce a locally symmetric state, we initialize the gauge degrees of freedom
in the product state \(\ket{\Omega} = \ket{1,\ldots, 1}\),
where~\(1\)~denotes the trivial element of \(G\),
and apply \(P\) to the state \(\ket{\psi}\otimes \ket{\Omega}\).
To summarize, the gauging procedure implements the map
\begin{equation}
  \label{eq:56}
  \ket{\psi} \longmapsto P(\ket{\psi}\otimes \ket{\Omega}) \eqcolon \mathcal{G}\ket{\psi}
\end{equation}
and is designed in such a way that the \emph{gauged state} \(\mathcal{G} \ket{\psi}\) satisfies
\begin{equation}
  \label{eq:57}
  \left(R(g) \otimes U_i(g) \otimes L(g)\right) \mathcal{G} \ket{\psi}= \mathcal{G} \ket{\psi}
\end{equation}
for all \(i\in \mathbb{Z}_n\).
We do not discuss how the gauging procedure applies to symmetric operators,
because they are not relevant for the present paper.

\subsection{Emergent \texorpdfstring{\(\Rep G\)}{Rep G} symmetry}
\label{sec:emergent-rep-g}
In addition to the local symmetry arising by construction,
the gauged state has a (so-called \emph{emergent}) global \(\Rep G\) symmetry \cite{haegemanGaugingQuantumStates2015, thorngrenFusionCategorySymmetry2024, lootensDualitiesOneDimensionalQuantum2023, garre-rubioEmergent2+1DTopological2024, blanikGaugingQuantumPhases2025},
acting on the gauge degrees of freedom:
\begin{equation}
  \mathcal{O}_{\rho}\,\mathcal{G} \ket{\psi} = \mathcal{G} \ket{\psi},
\end{equation}
where \(\rho\) is a representation of the group \(G\).
The action is given by 
\begin{equation}
  \label{eq:45}
  \mathcal{O}_{\rho}:
  \ket{g_1, \ldots, g_n}
  \longmapsto
  \frac{1}{d_{\rho}}\chi_\rho(g_1 \cdots g_n)
  \ket{g_1, \ldots, g_n},
\end{equation}
where \(d_{\rho}\) is the dimension of $\rho$
and \(\chi_\rho=\operatorname{tr}\circ \operatorname{\rho}\) is its character,
and it measures the flux labeled by \(\chi_\rho\).
To see that this indeed defines a symmetry on the gauged state,
notice that \(\mathcal{O}_{\rho}\) commutes with any \(\widehat{U}_i(g)\)
and thus with each local projector \(P_i\)
and acts as the identity on the auxiliary state \(\ket{\Omega}\).

Note that \(\mathcal{O}\) cannot constitute a group symmetry,
since for \(\sigma_1, \sigma_2 \in \operatorname{Irr}G\), 
the set of (representatives of) irreducible representations  of \(G\),
we have
\begin{equation}
  \label{eq:58}
  \mathcal{O}_{\sigma_1}\circ\, \mathcal{O}_{\sigma_2}
  = \mathcal{O}_{\sigma_1\otimes \sigma_2}
  = \sum_{\sigma\in \operatorname{Irr}G} N_{\sigma_1\sigma_2}^{\sigma} \mathcal{O}_{\sigma},
\end{equation}
where \(N_{\sigma_1\sigma_2}^{\sigma}\) denotes the multiplicity
of \(\sigma\) in the tensor product representation \(\sigma_1\otimes \sigma_2\).
In particular, most of the symmetry operators do not have an inverse of the form \(\mathcal{O}_{\rho}\),
which is the reason one often speaks of non-invertible symmetries in this context.
Because gauging implements a duality \cite{cuiperGaugingDualityOnedimensional2025, lootensDualitiesOneDimensionalQuantum2023, lootensDualitiesOnedimensionalQuantum2024},
the symmetry \(\mathcal{O}\) is considered to be \emph{dual} to the symmetry \(U\).
In fact, it is known that all duality operators can be written as gauging maps,
after disentangling the matter degrees of freedom from the gauge fields~\cite{lootensDualitiesOnedimensionalQuantum2024, cuiperGaugingDualityOnedimensional2025}.

To make contact with the general theory introduced below,
we mention that the operator \(\mathcal{O}_{\rho}\) can be written as
the MPO with periodic boundary conditions generated\footnote{
  The MPO has to be normalized to match Equation~\eqref{eq:45}.
}
from the tensors  
\begin{equation}
  \label{eq:46}
  
  \pgfkeys{
    /TNs/MPS/.cd,
    label inner = 1,
    label xpos = .2cm,
    index left = 1,
    index right = 1,
    index top = 1,
    index bottom = 1,
    label inner = \rho
  }
  \input{figures/MPORepTensor}

  \coloneq \sum_{g\in G}
  \mStk[1.5em]{{
  \pgfkeys{
    /TNs/dyad/.cd,
    label=\(\), 
    width=.5cm, 
    height=1.3cm, 
    pivot=1,
    element=g,
    
  }
  
\begin{tikzpicture}[baseline={([yshift=-\the\fontdimen22\textfont2]D.west)}, rotate=-90, transform shape] 
\if1\RotatePoint
  \node[empty] (D) at (0, 0) {\scriptsize\(\Ket{\rotatebox{90}{\(\el\)}}\)};
\else
  \node[empty] (D) at (0, 0) {\scriptsize\(\Ket{\rotatebox[origin = \RotatePoint]{90}{\(\el\)}}\)};
\fi
\node[empty] (Hr) at (-\heightTN/3, 0) {\TNLabel};
\draw[mid arrow] ([xshift=+\the\fontdimen10\textfont3]D.west) -- (Hr);
\node[empty] (lb) at (0, -\widthTN/2) {};
\node[empty] (rb) at (0,  \widthTN/2) {};
\end{tikzpicture}

}
             {
  \pgfkeys{
    /TNs/maps/.cd,
    label in=\(\), 
    label out=\(\), 
    width=1cm, 
    color=black,
    rev=0,
    arrows=mid arrow,
    map 1=1,
    map 2=1,
    map 3=1,
    pos 1=below,
    pos 2=below,
    pos 3=below,
    map 1 = \(\rho(g)\), color=black
  }
  \input{figures/tn-map-hz}
}
             {\raisebox{-5pt}{\(
  \pgfkeys{
    /TNs/dyad/.cd,
    label=\(\), 
    width=.5cm, 
    height=1.3cm, 
    pivot=1,
    element=g,
    
  }
  
\begin{tikzpicture}[baseline={([yshift=-\the\fontdimen22\textfont2]D.center)}, rotate=-90, transform shape] 
\if1\RotatePoint
  \node[empty] (D) at (0, 0) {\scriptsize\(\Bra{\rotatebox{90}{\(\el\)}}\)};
\else
  \node[empty] (D) at (0, 0) {\scriptsize\(\Bra{\rotatebox[origin = \RotatePoint]{90}{\(\el\)}}\)};
\fi
\node[empty] (Hr) at (\heightTN/3, 0) {\TNLabel};
\draw[mid arrow] (Hr) -- ([xshift=-\the\fontdimen10\textfont3]D.east);
\node[empty] (lb) at (0, -\widthTN/2) {};
\node[empty] (rb) at (0,  \widthTN/2) {};
\end{tikzpicture}

\)}}},
\end{equation}
where we use \({
  \pgfkeys{
    /TNs/dyad/.cd,
    label=\(\), 
    width=1.5cm, 
    element=g,
    
  }
  
\begin{tikzpicture}[baseline={([yshift=-\the\fontdimen22\textfont2]D.center)}] 
\node[empty] (D) at (0, 0) {\(\ket{\el}\)};
\node[empty] (Hl) at (-\widthTN/3, 0) {\TNLabel};
\draw [mid arrow] ([xshift=+\the\fontdimen10\textfont3]D.west) -- (Hl);
\end{tikzpicture}

\colon \mathbb{C}\rightarrow\mathbb{C}G}\) and
\({
  \pgfkeys{
    /TNs/dyad/.cd,
    label=\(\), 
    width=1.5cm, 
    element=h,
    element = g
  }
  
\begin{tikzpicture}[baseline={([yshift=-\the\fontdimen22\textfont2]D.center)}] 
\node[empty] (D) at (0, 0) {\(\bra{\el}\)};
\node[empty] (Hr) at (\widthTN/3, 0) {\TNLabel};
\draw [mid arrow] (Hr) -- ([xshift=-\the\fontdimen10\textfont3]D.east);
\end{tikzpicture}

\colon \mathbb{C}G\rightarrow\mathbb{C}}\)
to depict the canonical basis and dual basis vectors of
\(\mathbb{C}G\) and \(\mathbb{C}G^{\star}\), respectively. We note that the virtual level has dimension $d_\rho$. From the perspective of dualities, it is natural to ask whether it is possible to also gauge this
emergent dual symmetry and end up again with an on-site symmetry of the group \(G\).
The answer to this question turns out to be \emph{yes},
however, we require some more machinery to establish this precisely.

\subsection{Gauging MPO representations of categorical symmetries}
\label{sec:gauging-C}

The \(G\) and \(\Rep G\) symmetries just encountered both fit into the broader framework of
generalized finite symmetries described by unitary fusion categories (UFCs),
where a generalization to the standard gauging procedure
and its connection to dualities has recently been established~\cite{cuiperGaugingDualityOnedimensional2025}.

Let \(\mathcal{C}\) be a UFC and \(\operatorname{Irr} \mathcal{C}\)
a set of representatives of the isomorphism classes of simple objects in \(\mathcal{C}\).
Fix a basis of \(\operatorname{Hom}(a\otimes b, c)\) and the corresponding dual basis of
\(\operatorname{Hom}(c, a\otimes b)\), for each \(a,b,c\in \operatorname{Irr} \mathcal{C}\),
whose elements we depict as
\begin{align}
  \label{eq:67}
  \input{figures/StringBasis},&&
  \input{figures/StringDualBasis},
\end{align}
respectively, for \(i,j=1,\ldots, N^c_{ab}\),
where \(N^c_{ab}\) denotes the multiplicity of \(c\) in \(a\otimes b\).
These basis tensors can be seen as a generalization of the Clebsch-Gordan coefficients.
In this setting we consider an MPO representation of \(\mathcal{C}\)
to be a collection of MPO tensors
\begin{equation}
  \label{eq:27}
  
  \pgfkeys{
    /TNs/MPS/.cd,
    label inner = 1,
    label xpos = .2cm,
    index left = 1,
    index right = 1,
    index top = 1,
    index bottom = 1,
    index top = \mathcal{H}, index bottom = \mathcal{H}, index left = C, index right = C
  }
  \input{figures/MPORepTensor}
,
  \quad C\in \operatorname{Ob}\mathcal{C},
\end{equation}
labeled\footnote{
  We denote generic objects by capital letters and use lowercase letters only for simple objects.
}
by objects in \(\mathcal{C}\) with fixed Hilbert space \(\mathcal{H}\),
that are compatible with decomposition into a direct sum of simple objects
and satisfy
\begin{equation}
  \label{eq:53}
  \input{figures/SimpleStack}
  =\sum_{c\in \operatorname{Irr} \mathcal{C}}\sum_{i=1}^{N_{ab}^c}
  \input{figures/SimpleDecomp}
\end{equation}
for all \(a,b\in \operatorname{Irr} \mathcal{C}\).
Because all tensors we consider in the examples below can be viewed as linear maps,
we do not go into any more mathematical details here.
Through the MPO construction (with periodic boundaries)
one generates a collection of global symmetry operators \(\mathcal{O}_C\),
\begin{equation}
  \label{eq:22}
  \mathcal{O}_C\coloneq \frac{1}{\operatorname{dim}C}\input{figures/FMPO},
\end{equation}
labeled by objects in \(\mathcal{C}\)
on the Hilbert space \(\mathcal{H}^{\otimes n}\) 
that satisfy the fusion rules of \(\mathcal{C}\).
A state \(\ket{\psi}\) that satisfies \(\mathcal{O}_C \ket{\psi} = \ket{\psi} \)
for all objects \(C\) in \(\mathcal{C}\) is said to be \(\mathcal{C}\)-symmetric.

It is known~\cite{ostrikModuleCategoriesWeak2003, fuchsTFTConstructionRCFT2002, schaumannTracesModuleCategories2013}
that the data required to gauge a \(\mathcal{C}\)-symmetric state in \(1+1\) dimensions
can be encoded in the choice of a haploid, symmetric, special Frobenius algebra object
\(F \in \operatorname{Ob} \mathcal{C}\).
\begin{definition}[Frobenius algebra]
  A \emph{Frobenius algebra} internal to a monoidal\footnote{
    Very informally, a category where objects can be combined
    by a tensor product \(\otimes\) with a unit \(\mathbbm{1}\), coherently.
  }
  category~\(\mathcal{C}\) is an object~\(F\) in \(\mathcal{C}\) together with morphisms
  \begin{align}
    \label{eq:25}
    \eqb{25.1}{\mu \colon F\otimes F\longrightarrow F,} &&
    \eqb{25.2}{\Delta \colon F\longrightarrow F\otimes F,}
  \end{align}
  termed multiplication and comultiplication, respectively, and
  \begin{align}
    \label{eq:41}
    \eqb{25.1}{\eta\colon \mathbbm{1}\longrightarrow F,}&&
    \eqb{25.2}{\varepsilon\colon F\longrightarrow \mathbbm{1},}
  \end{align}
  termed unit and counit, respectively,
  constituting algebra and coalgebra structures on \(F\)
  that satisfy a compatibility condition that we give explicitly in Appendix~\ref{sec:frob-algebras}.
\end{definition}
The definitions of \emph{haploid}, \emph{symmetric} and \emph{special},
having no direct relevance to the present paper,
have also been deferred to Appendix~\ref{sec:frob-algebras}.

The proposed gauging procedure~\cite{cuiperGaugingDualityOnedimensional2025}
proceeds completely analogous to the group case in Subsection~\ref{sec:gauging-group}.
The Frobenius algebra \(F\) serves as the new gauge degrees of freedom after gauging
and the local projectors to the locally invariant subspaces,
generalizing the projectors in Equation \eqref{eq:47},
take the form
\begin{equation}
  \label{eq:60}
  P_i \coloneq \input{figures/tn-projector}
  \in \operatorname{End}(F\otimes \mathcal{H}\otimes F),
\end{equation}
constructed from the tensor \eqref{eq:27} for \(C = F\) and the product and coproduct of \(F\),
which we depict as
\begin{align}
  \label{eq:24}
  \mu \equiv \input{figures/StringDiagramMultiplication},&&
  \Delta \equiv \input{figures/StringDiagramCoMultiplication}.
\end{align}
In all examples we consider below, it is possible to endow \(F\) with a Hilbert space structure
and view \(\eta\) as a state in \(F\), hence we will use Hilbert space terminology going forward.
The operator implementing the gauging procedure,
i.e.\ initializing the auxiliary state as \(\ket{\Omega} = \bigotimes^n \eta \)
and projecting the product state to the locally invariant subspace of the global Hilbert space,
\(\ket{\psi}\mapsto \mathcal{G}_F\ket{\psi}\)
can be written as the following MPO
\begin{equation}
  \label{eq:23}
  \mathcal{G}_F \coloneq \input{figures/gaugingMPO}.
\end{equation}

\subsection{Twisted and partial gauging of group symmetries}
\label{sec:twist-part-gaug}
Global symmetries of finite groups, acting on a multi-particle Hilbert space through local representations,
are to be regarded as \(\vect^G\) symmetries in the framework of categorical symmetries,
where \(\vect^G\) denotes the category of finite dimensional \(G\)-graded vector spaces.
Note that the symmetry operator \(U(g)\) is of the form~\eqref{eq:27},
with \(C = \mathbb{C}\), viewed as a graded vector space of degree \(g\in G\).

It is well known~\cite[][Corollary 5.8]{turaevHomotopyQuantumField2010}
that (up to isomorphism) haploid, special, symmetric Frobenius algebras internal to \(\vect^G\)
are classified by tuples~\((K, \alpha)\), where \(K \leq G\) and
\(\alpha \in \operatorname{H}^2(K, \operatorname{U}(1))\).
The twisted group algebras \(\mathbb{C}^{\alpha}K\)
are the obvious representatives of the isomorphism classes.
In more detail, on the graded vector space \(\mathbb{C}K\)
take the Frobenius algebra structure given by
\begin{subequations}\label{eq:43}
\begin{align}
  \mu \vphantom{\frac{1}{|K|}}
  & \colon\eqb{43}{\ket{k}}\otimes \ket{h}\longmapsto\alpha(k,h)\ket{kh},\\
  \eta
  & \colon \eqb{43}{1} \longmapsto \ket{e},\\
  \Delta
  & \colon\eqb{43}{\ket{k}} \longmapsto \frac{1}{|K|}\sum_{h\in H}\frac{\ket{kh^{-1}}\otimes \ket{h}}{\alpha(kh^{-1}, h)},\\
  \varepsilon
  & \colon\eqb{43}{\ket{k}}\longmapsto|K|\,\delta_{k,e}.
\end{align}
\end{subequations}
This choice of Frobenius algebra as the input to the procedure outlined in Subsection \ref{sec:gauging-C},
with
\begin{equation}\label{eq:7}
  \eqmakebox[eq31]{\(
  \pgfkeys{
    /TNs/MPS/.cd,
    label inner = 1,
    label xpos = .2cm,
    index left = 1,
    index right = 1,
    index top = 1,
    index bottom = 1,
    label inner= \mathbb{C}^{\alpha}K, label xpos = -.5cm
  }
  \input{figures/MPORepTensor}
\)}
  = \sum_{k\in K} 
  \pgfkeys{
    /TNs/dyad/.cd,
    label=\(\), 
    width=1.5cm, 
    element=g,
    element = k
  }
  
\begin{tikzpicture}[baseline={([yshift=-\the\fontdimen22\textfont2]D.center)}] 
\node[empty] (D) at (0, 0) {\(\ket{\el}\)};
\node[empty] (Hl) at (-\widthTN/3, 0) {\TNLabel};
\draw [mid arrow] ([xshift=+\the\fontdimen10\textfont3]D.west) -- (Hl);
\end{tikzpicture}

\; 
  \pgfkeys{
    /TNs/maps/.cd,
    label in=\(\), 
    label out=\(\), 
    height=.8cm, 
    width=.4cm, 
    rev x=-1,
    rev y=-1,
    corners=0,
    map 1=1,
    pos 1=right,
    map 2=1,
    pos 2=right,
    height = 1cm, map 1 = \(U(k^{-1})\)
  }
  \input{figures/tn-map-v}

  \pgfkeys{
    /TNs/dyad/.cd,
    label=\(\), 
    width=1.5cm, 
    element=h,
    element = k
  }
  
\begin{tikzpicture}[baseline={([yshift=-\the\fontdimen22\textfont2]D.center)}] 
\node[empty] (D) at (0, 0) {\(\bra{\el}\)};
\node[empty] (Hr) at (\widthTN/3, 0) {\TNLabel};
\draw [mid arrow] (Hr) -- ([xshift=-\the\fontdimen10\textfont3]D.east);
\end{tikzpicture}

,
\end{equation}
perfectly reproduces the twisted gauging procedures, 
also labeled by \((K\leq G, \alpha\in \operatorname{H}^2(K, U(1)))\)
as defined, e.g., in~\cite{blanikGaugingQuantumPhases2025},
where the choice of subgroup \(K\leq G\) amounts to viewing the global \(G\) symmetry as a \(K\) symmetry.

Without the inverse \(k^{-1}\), appearing in Equation~\eqref{eq:7},
one would need a slightly different choice of Frobenius algebra than the one we give in \eqref{eq:43}
to describe gauging of group symmetries.
However, in the construction of the gauging map \(\mathcal{G}_{\mathbb{C}K}\),
cf.~Equation~\eqref{eq:23},
one can remove the inverse by simultaneously replacing \(\Delta\) with
\begin{equation}\label{eq:42}
  \Delta^{\prime} \coloneq
  \frac{1}{|K|} \sum_{k \in K}
  \Stk{
  \pgfkeys{
    /TNs/dyad/.cd,
    label=\(\), 
    width=.5cm, 
    height=1.3cm, 
    pivot=1,
    element=g,
    element = k
  }
  
\begin{tikzpicture}[baseline={([yshift=-\the\fontdimen22\textfont2]D.west)}, rotate=-90, transform shape] 
\if1\RotatePoint
  \node[empty] (D) at (0, 0) {\scriptsize\(\Ket{\rotatebox{90}{\(\el\)}}\)};
\else
  \node[empty] (D) at (0, 0) {\scriptsize\(\Ket{\rotatebox[origin = \RotatePoint]{90}{\(\el\)}}\)};
\fi
\node[empty] (Hr) at (-\heightTN/3, 0) {\TNLabel};
\draw[mid arrow] ([xshift=+\the\fontdimen10\textfont3]D.west) -- (Hr);
\node[empty] (lb) at (0, -\widthTN/2) {};
\node[empty] (rb) at (0,  \widthTN/2) {};
\end{tikzpicture}

}
      {
  \pgfkeys{
    /TNs/maps/.cd,
    label in=\(\), 
    label out=\(\), 
    width=1cm, 
    color=black,
    rev=0,
    arrows=mid arrow,
    map 1=1,
    map 2=1,
    map 3=1,
    pos 1=below,
    pos 2=below,
    pos 3=below,
    map 1 = \(\tw{\tau}{L}(k)\)
  }
  \input{figures/tn-map-hz}
},
\end{equation}
as we shall do below.
Note that while \(\Delta^{\prime}\) does not define a coproduct,
the replacement above just amounts to a change of basis on the virtual Hilbert space of the gauging MPO.
The left regular projective representation \(\tw{\tau}L\) of \(K\) appearing in Equation \eqref{eq:42}
is a generalization of the left regular representation, cf.~\eqref{eq:55},
and we take this opportunity to remind the reader
of the definitions of both it and the right regular projective representation,
which, as linear maps on \(\mathbb{C}K\), are given by
\begin{align} \label{eq:65}
  \tw{\tau}{L}(k) &\coloneq \sum_{h\in K} \tau(k, h) \dyad{kh}{h},\\
  \tw{\tau}{R}(k) &\coloneq \sum_{h\in K} \tau(h, k) \dyad{h}{hk}.
\end{align}

The gauging procedure reviewed in Section \ref{sec:gauging-group} is included as the case \((G, 1)\).
A non-trivial 2-cocycle is said to give rise to \emph{twisted} gauging
and the choice of subgroup \(K\leq G\) is interpreted to mean partial gauging.

\section{Gauging \texorpdfstring{\(\Rep G\)}{Rep G} symmetries}
\label{sec:gauging-rep-g}

In this section we discuss the gauging of \(\Rep G\) symmetries in more detail
and establish an explicit description of the dual \(G\) symmetry arising from the gauging procedure.
We consider \(\Rep G\) symmetries acting on the local Hilbert space \(\mathcal{H} = \mathbb{C}G\),
represented by MPOs of the form 
\begin{equation}
  \label{eq:19}
  
  \pgfkeys{
    /TNs/MPS/.cd,
    label inner = 1,
    label xpos = .2cm,
    index left = 1,
    index right = 1,
    index top = 1,
    index bottom = 1,
    label inner = \rho
  }
  \input{figures/MPORepTensor}

  \coloneq \sum_{g\in G}
  \mStk[1.5em]{{
  \pgfkeys{
    /TNs/dyad/.cd,
    label=\(\), 
    width=.5cm, 
    height=1.3cm, 
    pivot=1,
    element=g,
    
  }
  
\begin{tikzpicture}[baseline={([yshift=-\the\fontdimen22\textfont2]D.west)}, rotate=-90, transform shape] 
\if1\RotatePoint
  \node[empty] (D) at (0, 0) {\scriptsize\(\Ket{\rotatebox{90}{\(\el\)}}\)};
\else
  \node[empty] (D) at (0, 0) {\scriptsize\(\Ket{\rotatebox[origin = \RotatePoint]{90}{\(\el\)}}\)};
\fi
\node[empty] (Hr) at (-\heightTN/3, 0) {\TNLabel};
\draw[mid arrow] ([xshift=+\the\fontdimen10\textfont3]D.west) -- (Hr);
\node[empty] (lb) at (0, -\widthTN/2) {};
\node[empty] (rb) at (0,  \widthTN/2) {};
\end{tikzpicture}

}
               {
  \pgfkeys{
    /TNs/maps/.cd,
    label in=\(\), 
    label out=\(\), 
    width=1cm, 
    color=black,
    rev=0,
    arrows=mid arrow,
    map 1=1,
    map 2=1,
    map 3=1,
    pos 1=below,
    pos 2=below,
    pos 3=below,
    map 1 = \(\rho(g)\), color=black
  }
  \input{figures/tn-map-hz}
}
               {\raisebox{-5pt}{\(
  \pgfkeys{
    /TNs/dyad/.cd,
    label=\(\), 
    width=.5cm, 
    height=1.3cm, 
    pivot=1,
    element=g,
    
  }
  
\begin{tikzpicture}[baseline={([yshift=-\the\fontdimen22\textfont2]D.center)}, rotate=-90, transform shape] 
\if1\RotatePoint
  \node[empty] (D) at (0, 0) {\scriptsize\(\Bra{\rotatebox{90}{\(\el\)}}\)};
\else
  \node[empty] (D) at (0, 0) {\scriptsize\(\Bra{\rotatebox[origin = \RotatePoint]{90}{\(\el\)}}\)};
\fi
\node[empty] (Hr) at (\heightTN/3, 0) {\TNLabel};
\draw[mid arrow] (Hr) -- ([xshift=-\the\fontdimen10\textfont3]D.east);
\node[empty] (lb) at (0, -\widthTN/2) {};
\node[empty] (rb) at (0,  \widthTN/2) {};
\end{tikzpicture}

\)}}},\quad \rho\in \Rep G.
\end{equation}
We make this restriction to simplify notation and exposition,
because it is the symmetry that arises from gauging group symmetries, cf.~Subsection~\ref{sec:emergent-rep-g},
and thus has the most relevance to this paper.

\subsection{Frobenius algebras in \texorpdfstring{\(\Rep G\)}{Rep G}}
\label{sec:frob-algebr-rep-g}

It is well known \cite{ostrikModuleCategoriesWeak2003, etingofSemisimpleGEquivariantSimple2017} that
Frobenius algebras internal to \(\Rep G\) are classified, up to Morita equivalence,
by tuples~\((K, \alpha)\), where \(K \leq G\) and
\(\alpha \in \operatorname{H}^2(K, \operatorname{U}(1))\),
just like in the \(\vect^G\) (group symmetry) case.
As explicit representatives
\cite{ostrikModuleCategoriesWeak2003, etingofSemisimpleGEquivariantSimple2017, kapustinTopologicalFieldTheory2017}
one can take, e.g., the representations\footnote{
  We use \(\operatorname{Ind}\) to denote the induced representation and
  \(\operatorname{End}V\) is the space of endomorphisms on \(V\) in \(\Vect\),
  together with the adjoint representation of \(K\).
}
\(\operatorname{Ind}^G_K \operatorname{End}V\),
where \((V,\rho)\in \Rep^{\alpha}_{\mathbb{C}}K\)
is irreducible,
with the obvious Frobenius algebra structure.
In more detail, take \(F\) to be the space of functions~\(f\colon G\rightarrow \operatorname{End}V\),
satisfying 
\begin{equation}
  \label{eq:16}
  f(gk)=\rho(k)^{-1}f(g)\rho(k)
\end{equation}
for all \(g\in G\) and \(k\in K\),
and let \(G\) act on it through left translation,
i.e.\ \((l_{g} f)(g^{\prime}) = f(g^{-1}g^{\prime})\).
The algebra structure on \(F\) is given by pointwise multiplication
and the Frobenius trace is
\begin{equation}
  \label{eq:28}
  f \longmapsto \sum_{g\in G} \operatorname{tr} f(g),
\end{equation}
where \(\operatorname{tr}\) is just the trace in \(\operatorname{End}V\).
Both of these operations are clearly \(G\)-equivariant,
and it is clear that the Frobenius algebras thus constructed are symmetric.
Since the multiplicity of the trivial representation in \(\operatorname{End}V\)
is equal to one by Schur's lemma and invariant under induction by Frobenius reciprocity,
\(F\) is haploid.
By choosing a basis of \(F\) and explicitly constructing \(\Delta\), it is easy to show that \(F\) is special.

\subsection{Dual symmetry}
\label{sec:dual-symmetry}

In the case where \(K\trianglelefteq G\) is a normal subgroup and
the representation \(V\) is chosen to be trivial,
one can also define right translation \((r_{g} f)(g^{\prime}) \coloneq f(g^{\prime}g)\) on
the Frobenius algebra \(F\), effectively\footnote{
  Note that \(r\) restricted to \(K\) acts trivially on \(F\).
}
giving a representation of \(G/K\) on \(F\).
Since multiplication \(\mu\colon F\otimes F\rightarrow F\) and Frobenius trace
are also equivariant under right translation
and the coproduct \(\Delta\colon F\rightarrow F\otimes F\)
is just the dual of the multiplication under the identification \(F\cong F^{\star}\)
induced by the trace,
the coproduct is also equivariant with respect to \(r\):
\begin{equation}
  \label{eq:38}
  \Delta \circ \eqb{38.1}{r_g} = (\eqb{38.2}{r_g\otimes r_g}) \circ \Delta.
\end{equation}
Since \(r\) and \(l\) commute, it is clear from \eqref{eq:23} that
the gauging operator \(\mathcal{G}_F\) is invariant under the global action of \(r_g\)
on the newly introduced degrees of freedom
\begin{equation}
  \label{eq:52}
  \mathcal{G}_F = \left( \cdots \otimes r_g\otimes r_g\otimes r_g\otimes \cdots\right)
  \circ \mathcal{G}_F
\end{equation}
for all \(g \in G\).
The same action leaves any gauged state invariant 
and we have thus established the explicit form of an emergent dual \(G/K\)-symmetry.

Only for \(K= \left\{ 1 \right\}\) does a full \(G\) symmetry emerge after gauging.
Thus, it is natural to regard the choice \((K=\{1\}, \alpha=1)\) and \(V = \mathbb{C}\)
as corresponding to untwisted gauging of the full symmetry,
leading us to make the following definition.
\begin{definition}[Full untwisted gauging]\label{def:full-untwisted-gauging-repg}
  We will call the gauging procedure applied to a \(\Rep G\) symmetry, 
  with the choice of Frobenius algebra structure\footnote{
    Morphisms in \(\Rep G\) are intertwiners.
  } on the left regular representation given by 
  \begin{subequations}
  \begin{align}
    \label{eq:3}
    \mu
    &\colon \eqb{3}{\ket{g}}\otimes \ket{h}\longmapsto \delta_{g,h}\ket{g},\\
    \eta
    &\colon \eqb{3}{1}\longmapsto \sum_{g\in G}\ket{g},\\
    \Delta
    &\colon \eqb{3}{\ket{g}}\longmapsto \ket{g}\otimes \ket{g},\\
    \varepsilon
    &\colon \eqb{3}{\ket{g}}\longmapsto 1,
  \end{align}
  \end{subequations}
  \emph{full untwisted gauging}.
\end{definition}

From a more fundamental point of view,
if every simple object of the fusion category \(\mathcal{C}\) appears in the decomposition of \(F\)
into simple objects at least once,
then the gauging defined by \(F\) is called \emph{full},
and called \emph{partial} otherwise.
Generally, it can happen that a fusion category does not admit a full gauging procedure,
in which case the symmetry is called \emph{anomalous}.
Since the regular group representations decompose into all irreps,
Definition~\ref{def:full-untwisted-gauging-repg} agrees with this interpretation.

In the case of full untwisted gauging, it is immediate that the Frobenius algebra
\(\left( L, \mu, \eta, \Delta, \varepsilon \right)\)
is haploid, symmetric and special.
Graphically, we depict comultiplication as
\begin{equation}
  \label{eq:6}
  \Delta \equiv \input{figures/DefCoProduct} \coloneq\sum_{g\in G}\,\input{figures/DefDot},
\end{equation}
which is simply a delta tensor.
Substituting into Equation~\eqref{eq:23},
we obtain that the gauging MPO \(\mathcal{G}_L\),
associated to the Frobenius algebra object~\(L\in \Rep G\)
of Definition~\ref{def:full-untwisted-gauging-repg},
is given by
\begin{equation}
  \label{eq:21}
  \mathcal{G}_L = \input{figures/gaugingMPOex}.
\end{equation}
Here, the right regular representation \(R\) of \(G\) takes the role of the emergent dual symmetry
and Equation \eqref{eq:38} can be written as
\begin{equation}
  \label{eq:37}
  
\begin{tikzpicture}[baseline={([yshift=-\the\fontdimen22\textfont2]0.center)}, y={(1.0em,0.5em)},x={(1.0em,-0.5em)}, z={(0em,1em)}]

\node [matrix, fill=white] (0) at (0, 0) {\(\)};

\providecommand{\LocalWidth}{}
\providecommand{\LocalHeight}{}
\renewcommand{\LocalWidth}{3/2}
\renewcommand{\LocalHeight}{3/2}
\draw [mid arrow] (0)  -- (0, 0, \LocalHeight);
\draw [mid arrow] (\LocalWidth, 0)  -- (0);
\draw [mid arrow] (0) -- (0, -\LocalWidth);
\end{tikzpicture}
\;=\;
  \input{figures/ProdSym2}.
\end{equation}
To reiterate, since \([L(g), R(h)] = 0\) for all \(g,h\in G\),
we have the emergent global \(G\) symmetry
\begin{equation}
  \label{eq:9}
  R(g)\otimes \cdots \otimes R(g),
\end{equation}
acting on the newly introduced gauge degrees of freedom.
This can also be seen from the local projectors
\begin{equation}
  \label{eq:54}
  P_i = \sum_{g,h\in G}
  \dyad{g}{g}\otimes
  \dyad{gh^{-1}}{gh^{-1}}\otimes
  \dyad{h}{h},
\end{equation}
which are the explicit description of the projectors \eqref{eq:60}
for the case of full untwisted gauging of \(\Rep G\).
They can be viewed as being dual to the projectors we defined in \eqref{eq:47},
and there is a way~\cite{cuiperGaugingDualityOnedimensional2025} to view them as a sum of symmetry operators,
just like the projectors~\eqref{eq:47} in the group symmetric case.
Notice that \(R(g)\otimes \mathbbm{1}\otimes R(g)\) commutes with the local projector \(P_i\)
and leaves the auxiliary state invariant
\begin{equation}
  \label{eq:66}
  R(g)\circ \eta \equiv R(g)\sum_{h\in G}\ket{h}
  =\sum_{h\in G}\ket{h}
  \equiv \eta.
\end{equation}

\section{Emergent quantum doubles}
\label{sec:emerg-quant-doubl}

In this section we will discuss how ground states
of the quantum double model emerge
through an iterated application of the gauging procedures described above.
We begin by constructing a specific state \(\ket{\psi}\),
to be used as the input state to the gauging procedure.
Let \(K\leq G\) and \(\beta\in \operatorname{H}^2(K,\operatorname{U}(1))\),
and consider the 2-site translation invariant MPS
\begin{equation}
  \label{eq:31}
  \ket{\psi^{\prime}} \coloneq
  \cdots \input{figures/MPS-input}\cdots,
\end{equation}
where
\begin{equation}
  \label{eq:33}
  \input{figures/boundaryMPS}
  \coloneq 
  \sum_{k\in K}
  \mStk[1.8em]{{
  \pgfkeys{
    /TNs/dyad/.cd,
    label=\(\), 
    width=.5cm, 
    height=1.3cm, 
    pivot=1,
    element=g,
    element = k
  }
  
\begin{tikzpicture}[baseline={([yshift=-\the\fontdimen22\textfont2]D.west)}, rotate=-90, transform shape] 
\if1\RotatePoint
  \node[empty] (D) at (0, 0) {\scriptsize\(\Ket{\rotatebox{90}{\(\el\)}}\)};
\else
  \node[empty] (D) at (0, 0) {\scriptsize\(\Ket{\rotatebox[origin = \RotatePoint]{90}{\(\el\)}}\)};
\fi
\node[empty] (Hr) at (-\heightTN/3, 0) {\TNLabel};
\draw[mid arrow] ([xshift=+\the\fontdimen10\textfont3]D.west) -- (Hr);
\node[empty] (lb) at (0, -\widthTN/2) {};
\node[empty] (rb) at (0,  \widthTN/2) {};
\end{tikzpicture}

}
               {
  \pgfkeys{
    /TNs/maps/.cd,
    label in=\(\), 
    label out=\(\), 
    width=1cm, 
    color=black,
    rev=0,
    arrows=mid arrow,
    map 1=1,
    map 2=1,
    map 3=1,
    pos 1=below,
    pos 2=below,
    pos 3=below,
    map 1 = \(\tw{\beta}{L}(k)\)
  }
  \input{figures/tn-map-hz}
}
               {}}.
\end{equation}
The state \(\ket{\psi^{\prime}}\) has a local \(K\)~symmetry, represented by
\begin{equation}
  \label{eq:34}
  \cdots \otimes \mathbbm{1} \otimes
  R^{\beta}(k)\otimes L(k) \otimes L^{\overline{\beta}}(k)
  \otimes \mathbbm{1} \otimes \cdots,
\end{equation}
and a global \(K\) symmetry, represented by
\begin{equation}
  \label{eq:32}
  \widehat{R}(k)\coloneq
  \cdots \otimes 
  R(k)\otimes \mathbbm{1} \otimes R(k) \otimes \mathbbm{1}\otimes R(k)
  \otimes \cdots.
\end{equation}
Since we want a \(G\)-symmetric state,
we apply a symmetrization to \(\ket{\psi^{\prime}}\).
To that end, let \(r \coloneq |G/K|\) and choose \(g_1, \ldots, g_r\in G\) such that
\(G = \bigcup_{i=1}^r g_i K\).
The state
\begin{equation}
  \label{eq:71}
  \ket{\psi}\coloneq
  \sum_{i=1}^r \widehat{R}(g_i)\ket{\psi^{\prime}}
  \simeq \bigoplus_{i=1}^r \widehat{R}(g_i)\ket{\psi^{\prime}}
\end{equation}
has a global \(G\)-symmetry represented by \(\widehat{R}(g)\),
while retaining the local \(K\) symmetry \eqref{eq:32} of \(\ket{\psi^{\prime}}\)
which of course commutes with \(\widehat{R}\).
With respect to its global \(K\) symmetry,
the state \(\ket{\psi^{\prime}}\) is in the SPT phase
labeled by \((K=K, \alpha)\), hence, when \(K\) is normal in \(G\),
the state~\(\ket{\psi}\) will exhibit symmetry breaking
and belongs to the phase \((K\trianglelefteq G, \alpha)\),
with respect to its global \(G\) symmetry.

Note that, since \(\mathcal{G}_{\mathbb{C}^{\tau}G}\) is right absorbing
with respect to the global symmetry,
we have up to normalization
\begin{equation}
  \label{eq:73}
  \mathcal{G}_{\mathbb{C}^{\tau}G}\ket{\psi}
  = \mathcal{G}_{\mathbb{C}^{\tau}G}\ket{\psi^{\prime}}.
\end{equation}
When gauging the full \(G\) symmetry, we have a choice of 2-cocycle
\(\tau\in \operatorname{H}^2(G, \operatorname{U}(1))\),
cf.~Section~\ref{sec:twist-part-gaug},
and applying the \(\tau\)-twisted gauging procedure,
labeled by \(\mathbb{C}^{\tau}G\),
to the global \(G\)-symmetry yields a state with a global \(\Rep G\) symmetry.
Applying full untwisted gauging, labeled by \(L\),
to this symmetry in turn results in a state with a global \(G\)-symmetry.
Note that there is only a single choice for gauging the full \(\Rep G\) symmetry.
Iterating this procedure yields the state 
\begin{equation}
  \label{eq:72}
  \mathcal{G}\ket{\psi}\coloneq \cdots
  \circ\mathcal{G}_L
  \circ\mathcal{G}_{\mathbb{C}^{\tau}G}
  \circ\mathcal{G}_L
  \circ\mathcal{G}_{\mathbb{C}^{\tau}G}
  \ket{\psi}
\end{equation}
whose PEPS representation is depicted in Figure \ref{fig:peps-representation},
where we use the following PEPS tensors
\begin{align}
  
  \pgfkeys{
    /TNs/PEPS/.cd,
    width=2, 
    height=2, 
    label=\(\), 
    label inner=\(\),
    pattern = {},
    map top = 1,
    map LU = 1,
    map LD = 1,
    map RU = 1,
    map RD = 1,
    label = \(T_{\mathrm{o}}\), pattern=A
  }
  \input{figures/peps-tensor}
\,&\coloneq\sum_{g\in G}\,\input{figures/DefA},\\
  
  \pgfkeys{
    /TNs/PEPS/.cd,
    width=2, 
    height=2, 
    label=\(\), 
    label inner=\(\),
    pattern = {},
    map top = 1,
    map LU = 1,
    map LD = 1,
    map RU = 1,
    map RD = 1,
    label = \(\tw{\tau}{T}_{\mathrm{e}}\), pattern=B
  }
  \input{figures/peps-tensor}
\,&\coloneq\sum_{g\in G}\,\input{figures/DefB}.
\end{align}
\begin{figure}[H]
  \centering
  \input{figures/PEPS-bdry}
  \caption{
    PEPS representation of the state \(\mathcal{G}\ket{\psi}\),
    obtained through the iterated gauging procedure detailed above.
    The tensors \(\tw{\tau}{T}_{\mathrm{e}}\) and \(T_{\mathrm{o}}\) appear
    on even and odd layers of the PEPS representation, respectively,
    and the tensors \(\tw{\beta}{T}_{\mathrm{b}}\)
    constitute the bottom boundary.
  }
  \label{fig:peps-representation}
\end{figure}
It is convenient to view the state \(\mathcal{G}\ket{\psi}\)
as having degrees of freedom living on the edges of an oriented square lattice,
as depicted in Figure \ref{fig:square-lattice}.
The purpose of the orientation,
which is going from right to left and top to bottom,
will become apparent once we make the connection to quantum double models.
\begin{figure}[H]
  \centering
  \input{figures/PEPS-sqr}
  \caption{
    Degrees of freedom on the edges of a square lattice
    with smooth bottom boundary.
    The orientation of the edges is used in the definition of the Hamiltonian.
    We have indicated schematically both the action of the vertex operators
    \(S_{v_1}(k)\) and \(S_{v_2}(g)\),
    as well as the action of the plaquette operators
    \(P_{f_1}(\sigma)\) and \(P_{f_2}(\rho)\).
  }
  \label{fig:square-lattice}
\end{figure}
Every time we gauge a group symmetry, we could use a different choice of 2-cocycle,
leading to non-abelian generalizations of the models studied in~\cite{rubioDipolesAnyonicDirectional2025}.
Because this would not affect the following arguments substantially,
we opt to keep the notation simple by always using the same 2-cocycle when gauging group symmetries.

On the square lattice we define star operators~\(S_v\),
labeled by vertices~\(v\),
and plaquette operators~\(P_f\),
labeled by faces~\(f\).
In the bulk, the star operators form representations of \(G\) and are given by
\begin{equation}
  \label{eq:8}
  S_v(g)
  \coloneq 
\begin{tikzpicture}[baseline={([yshift=-\the\fontdimen22\textfont2]0.center)}]

\coordinate (0) at (0, 0) {};

\node [color=lightgray, tensor] (L) at (-\scale/2, 0) {\(\)};
\node [color=lightgray, tensor] (R) at (\scale/2, 0) {};
\node [color=lightgray, tensor] (T) at (0, \scale/2) {};
\node [color=lightgray, tensor] (B) at (0, -\scale/2) {};
\draw [color=lightgray, offset arrow] (R) -- (L);
\draw [color=lightgray, offset arrow] (T) -- (B);
\draw [color=lightgray](T) --++ (0,1/2);
\draw [color=lightgray](B) --++ (0,-1/2);
\draw [color=lightgray](R) --++ (1/2, 0);
\draw [color=lightgray](L) --++ (-1/2, 0);
\node [color=gray]at (-1/4, -1/4) {\scriptsize\(v\)};
\node at (L) {\VertexOperatorLeft};
\node at (R) {\VertexOperatorRight};
\node at (T) {\VertexOperatorTop};
\node at (B) {\VertexOperatorBottom};

\end{tikzpicture}
\,
  \equiv \;\input{figures/StarOperator3D},
\end{equation}
for \(g\in G\), while on the boundary they form representations of \(K\) and are given by
\begin{equation}
  \label{eq:35}
  S_v(k)
  \coloneq 
\providecommand{\scale}{4.3}
\renewcommand{\scale}{3}
\begin{tikzpicture}[baseline={([yshift=-\the\fontdimen22\textfont2]0.center)}]

\coordinate (0) at (0, 0) {};

\node [color=lightgray, tensor] (L) at (-\scale/2, 0) {\(\)};
\node [color=lightgray, tensor] (R) at (\scale/2, 0) {};
\node [color=lightgray, tensor] (T) at (0, \scale/2) {};
\draw [color=lightgray, offset arrow] (R) -- (L);
\draw [color=lightgray, mid arrow] (T) -- (0);
\draw [color=lightgray](T) --++ (0,1/2);
\draw [color=lightgray](R) --++ (1/2, 0);
\draw [color=lightgray](L) --++ (-1/2, 0);
\node [color=gray]at (0, -1/4) {\scriptsize\(v\)};
\node at (L) {\OperatorLeft};
\node at (R) {\OperatorRight};
\node at (T) {\OperatorTop};

\end{tikzpicture}
\,
  \equiv\; \input{figures/StarOperatorBdry3D},
\end{equation}
for \(k\in K\).
The plaquette operators in the bulk form a representation of \(\Rep G\),
while those on the boundary form a representation of \(\Rep K\),
but, since the boundary of the square lattice is smooth,
they all can be written as
\begin{align}
  \label{eq:17}
  P_f(\rho)
  \coloneq \frac{1}{d_{\rho}}\input{figures/PlaquetteOperator}
  \equiv \frac{1}{d_{\rho}}\input{figures/PlaquetteOperator3D},
\end{align}
where \(\rho\) belongs to \(\Rep G\) or \(\Rep K\)
and the sum in Equation~\eqref{eq:19} runs over \(G\) or \(K\), respectively.
We use color to aid visualization. It has no other significance.
The definition of \(P\) must not be interpreted as a string diagram in a dagger category,
but should rather be read as
\begin{equation}
  \label{eq:10}
  P_f(\rho) \sim \sum \operatorname{tr}
  \left[\rho(g)\rho(h)\rho(k)^{\dag}\rho(l)^{\dag}\right],
\end{equation}
which measures the flux through the plaquette \(f\).
Notably, the inverse group element is evaluated in the definition of \(P\),
wherever the orientation of \(P\) and the orientation of the square lattice are antialigned.

As operators on the multi-particle Hilbert space we have
for all vertices \(v\neq v^{\prime}\) and faces \(f\) and \(f^{\prime}\),
\begin{align} \label{eq:11}
  \left[S_v, S_{v^{\prime}}\right]
  = \left[S_v, P_f\right]
  = \left[P_f, P_{f^{\prime}}\right]
  = 0.
\end{align}
The last equality is trivial, since the plaquette operators are diagonal in the canonical basis,
and the two other equalities are easy to compute explicitly.
To show that the state \(\mathcal{G}\ket{\psi}\) is invariant under
star and plaquette operators, we compute
\begin{subequations}
  \label{eq:48}
\begin{align}
  \input{figures/StarProof1}\!
  \stackrel{\eqref{eq:63}}{=}\,&\input{figures/StarProof2}\\
  \stackrel{\phantom{\eqref{eq:63}}}{=}\,&\input{figures/StarProof3},\\
  \input{figures/PlaquetteProof1}
  \stackrel{\eqref{eq:64}}{=}\;&\input{figures/PlaquetteProof2}\\
  =d_{\rho}\;&\input{figures/PlaquetteProof3},
\end{align}
\end{subequations}
where
\begin{equation}
  \label{eq:18}
  \input{figures/RepGMPS}
  \coloneq \sum_{g\in G}
  \mStk[1.1em]{{}
               {
  \pgfkeys{
    /TNs/maps/.cd,
    label in=\(\), 
    label out=\(\), 
    width=1cm, 
    color=black,
    rev=0,
    arrows=mid arrow,
    map 1=1,
    map 2=1,
    map 3=1,
    pos 1=below,
    pos 2=below,
    pos 3=below,
    map 1 = \(\rho(g)\), pos 1 = above, color=violet
  }
  \input{figures/tn-map-hz}
}
               {
  \pgfkeys{
    /TNs/dyad/.cd,
    label=\(\), 
    width=.5cm, 
    height=1.3cm, 
    pivot=1,
    element=g,
    
  }
  
\begin{tikzpicture}[baseline={([yshift=-\the\fontdimen22\textfont2]D.center)}, rotate=-90, transform shape] 
\if1\RotatePoint
  \node[empty] (D) at (0, 0) {\scriptsize\(\Bra{\rotatebox{90}{\(\el\)}}\)};
\else
  \node[empty] (D) at (0, 0) {\scriptsize\(\Bra{\rotatebox[origin = \RotatePoint]{90}{\(\el\)}}\)};
\fi
\node[empty] (Hr) at (\heightTN/3, 0) {\TNLabel};
\draw[mid arrow] (Hr) -- ([xshift=-\the\fontdimen10\textfont3]D.east);
\node[empty] (lb) at (0, -\widthTN/2) {};
\node[empty] (rb) at (0,  \widthTN/2) {};
\end{tikzpicture}

}}.
\end{equation}
The computations for the boundary terms proceed along similar lines.
It is thus clear that \(\mathcal{G}\ket{\psi}\) is a ground state 
of the commuting projector Hamiltonian
\begin{equation}
  \label{eq:14}
  H \coloneq -\sum_v A_v -\sum_f B_f,
\end{equation}
where
\begin{align*}
  A_v &\coloneq
  \begin{dcases}
    \frac{1}{|G|}\sum_{g\in G} S_v(g), &v\in\text{bulk},\\
    \frac{1}{|K|}\sum_{k\in K} S_v(k), &v\in\text{boundary},
  \end{dcases}\\
  B_f &\coloneq
  \begin{dcases}
    \frac{1}{|\Rep G\,|}\sum_{\rho} P_f(\rho) &f\in\text{bulk},\\
    \frac{1}{|\Rep K\,|}\sum_{\sigma} P_f(\sigma) &f\in\text{boundary}.
  \end{dcases}
\end{align*}
Note that for \(\tau = 1\) the Hamiltonian \(H\) is precisely
the Hamiltonian of the quantum double model with boundary~\cite{beigiQuantumDoubleModel2011}
labeled by \(\alpha\).

Finally, we mention that the input state \(\ket{\psi^{\prime}}\)
can itself be viewed as resulting from \(\tau\)-twisted gauging
of the state \(\ket{+}\in \mathbb{C}K^{\otimes n}\), cf.~\cite{blanikGaugingQuantumPhases2025}.
Gauging its global \(\Rep K\) symmetry,
which can naturally be viewed as a \(\Rep G\) symmetry,
one can extend Figures~\ref{fig:peps-representation} and~\ref{fig:square-lattice} downwards,
thus eliminating the boundary.

\section{Gauging of \texorpdfstring{\(G\)}{G} and \texorpdfstring{\(\Rep G\)}{Rep G} symmetries in 2+1 dimensions}

In this section we review how to gauge a global group symmetry
of a state on a two-dimensional lattice. \pagebreak
We show that a 1-form\footnote{
  The symmetry action of a 1-form symmetry on a two-dimensional lattice
  is represented by string like objects.
}
\(\Rep G\) symmetry,
acting on the gauge degrees of freedom, emerges.
We then propose a procedure to gauge this 1-form symmetry
and establish that it produces a dual \(G\) symmetry again.
Finally, we show that iterative application of the gauging procedure
yields quantum double models in three spatial dimensions.
We do not consider subsystem symmetries other than the ones just mentioned.

\subsection{Gauging \texorpdfstring{\(G\)}{G} symmetries}
\label{sec:gauging-g-symmetries}

Consider a \(G\)-symmetric quantum state \(\ket{\psi}\)
on a two-dimensional square lattice.
We assume that the lattice is oriented and treat its
vertices and edges as constituting an oriented graph \(\Lambda = (V, E)\).
We remind the reader of the relevant notation in Appendix~\ref{sec:graphs}.
Hence, to each vertex \(v\in V = \operatorname{vert}\Lambda\)
we associate the (for simplicity, same) Hilbert space \(\mathcal{H}_v\),
together with a representation \((\mathcal{H}_v, U_v)\in \Rep G\).
The statement that \(\ket{\psi}\) is symmetric then becomes
\begin{equation}
  \label{eq:12}
  \bigotimes _{v\in V} U_v(g) \ket{\psi} = \ket{\psi}
  \quad \text{for all } g\in G.
\end{equation}
Mimicking the approach in one dimension,
we introduce new degrees of freedom associated to every edge
\(e\in E = \operatorname{edge}\Lambda\) of the lattice,
described by the Hilbert space \(\mathcal{H}_e \coloneq \mathbb{C}G\),
on which the left regular \(L_e = L\) and right regular \(R_e = R\)
representations of \(G\) act.
We define the projection to the symmetric subspace at
\(v\in V\)
to be
\begin{equation}
  \label{eq:13}
  P_v \coloneq \frac{1}{|G|}\sum_{g\in G} U_v(g)
  \bigotimes_{e\in E_v^{\mathrm{o}}}R_e(g)
  \bigotimes_{e\in E_v^{\mathrm{t}}}L_e(g).
\end{equation}
As the auxiliary state, we choose the product state made up
of neutral group elements,
which gives the following gauging map:
\begin{equation}
  \label{eq:15}
  \mathcal{G}_0\colon\ket{\psi}\longmapsto
  \left(\prod_{v\in V} P_v \right)
  \left(
    \ket{\psi} \bigotimes_{e\in E}\ket{1}_e
  \right).
\end{equation}
Incidentally, a global symmetry can be considered to be a 0-form symmetry,
hence the notation \(\mathcal{G}_0\).
As in the one-dimensional case,
a dual \(\Rep G\) symmetry emerges because the symmetrization procedure
imposes a zero flux condition along closed paths on the gauge degrees of freedom.
Let \(\gamma = (v_1, \ldots, v_n, v_1)\) be a closed path in \(\Lambda\)
and for \(i=1, \ldots, n\) define \(\gamma_i\) to be equal to \(1\)
if \(e_i\coloneq (v_i, v_{i+1})\in E\)
and equal to \(-1\) otherwise.\footnote{
  Note that \((v_i, v_{i+1})\notin E\) means the
  corresponding edges in \(\gamma\) and \(\Lambda\) have opposite orientations.
}
For any representation \(\rho\in \Rep G\) we define the following operator
\begin{equation}
  \label{eq:1formsym}
  \mathcal{O}_{\gamma}(\rho)\colon
  \bigotimes_{e\in E} \ket{g_e}_e \longmapsto
  \frac{1}{d_{\rho}}
  \chi_{\rho}(g_{e_1}^{\gamma_1}\cdots g_{e_n}^{\gamma_n})
  \bigotimes_{e\in E} \ket{g_e}_e,
\end{equation}
where \(\chi_{\rho}\coloneq \operatorname{tr}\circ \operatorname{\rho}\)
is the character of \(\rho\) and \(d_{\rho}\) is its degree.
Note that the group element corresponding to an edge in the path
appears with an inverse in the trace if the corresponding edge in the graph
has the opposite orientation.
We interpret \(\mathcal{O}_{\gamma}(\rho)\) to measure the flux of \(\rho\)
through the loop \(\gamma\).
Since \(\mathcal{O}_{\gamma}(\rho)\) commutes with each \(P_v\)
and leaves invariant the auxiliary state \(\bigotimes_{e\in E} \ket{1}_e\),
we can conclude that
\begin{equation}
  \label{eq:20}
  \mathcal{O}_{\gamma}(\rho)\circ \mathcal{G}_0 = \mathcal{G}_0.
\end{equation}

For an in depth study of 1-form symmetries in terms of topological surface operators and defects,
see~\cite{bartschNoninvertibleSymmetriesHigher2024,bhardwajUniversalNonInvertibleSymmetries2022,delcampHigherCategoricalSymmetries2024}. 

\subsection{Gauging a 1-form \texorpdfstring{\(\Rep G\)}{Rep G} symmetry}

To gauge a 1-form \(\Rep G\) symmetry,
we propose a prescription that is dual to the one in the previous section.
This has the benefit of being directly applicable to states produced
by gauging a \(G\)-symmetric state.
Hence, we now let the local degrees of freedom of a quantum state \(\ket{\varphi}\) 
be described by Hilbert spaces \(\mathcal{H}_e\),
associated to the edges of a square lattice, viewed as the graph \(\Lambda\).
We shall focus on the specific 1-form \(\Rep G\) symmetry
generated by the operators \(\mathcal{O}_{\gamma}(\rho)\)
introduced in the previous section,
where \(\rho\in \Rep G\) and \(\gamma\) is a closed path in \(\Lambda\). 

We begin by introducing new degrees of freedom on the vertices \(v\in V\)
of the lattice,
described by the Hilbert space \(\mathcal{H}_v \coloneq \mathbb{C}G\).
Notice that for a fixed path~\(\gamma\), the operators
\(\mathcal{O}_{\gamma}(\rho)\) for \(\rho\in \Rep G\),
generate a global \(\Rep G\) symmetry on the Hilbert space
\(\bigotimes_{\gamma} \mathcal{H}_e\) along the path.
Hence, we propose to use the same local projectors, cf.~Equation~\eqref{eq:54},
that we introduced in the setting of global \(\Rep G\) symmetries:
\begin{equation}
  \label{eq:68}
  P_e = \sum_{g,h\in G}
  \dyad{g}{g}_{\operatorname{t}(e)}\otimes
  \dyad{gh^{-1}}{gh^{-1}}_e\otimes
  \dyad{h}{h}_{\operatorname{o}(e)},
\end{equation}
all of which commute,
i.e.~\([P_e,P_{e^{\prime}}]=0\) for all \(e, e^{\prime}\in E\).
More generally, the categorical gauging procedure can easily
be extended to 1-form symmetries on a two-dimensional lattice,
provided the chosen Frobenius algebra is commutative.
Note that the Frobenius algebras encoding partial (non-twisted)
gauging of \(\Rep G\) are all commutative.

We choose as the auxiliary state the product of plus states
\(\ket{+}\coloneq \sum_g\ket{g}\)
and define the gauging procedure to be
\begin{equation}
  \label{eq:36}
  \mathcal{G}_1\colon\ket{\varphi}\longmapsto
  \left(\prod_{e\in E} P_e \right)
  \left(\ket{\varphi} \bigotimes_{v\in V}\ket{+}_v \right),
\end{equation}
for \(\Rep G\) invariant states \(\ket{\varphi}\),
i.e. \( \mathcal{O}_{\gamma}(\rho)\ket{\varphi} = \ket{\varphi}\)
for all closed paths \(\gamma\) and representations \(\rho\in \Rep G\).

Since \(R(g)\ket{+} =\ket{+}\) and
\(R_{\mathrm{t}(e)}(g)\otimes \mathbbm{1}_e\otimes R_{\mathrm{o}(e)}(g)\)
commutes with the projector \(P_e\),
the gauged state \(\mathcal{G}_1\ket{\varphi}\)
has an emergent dual global \(G\) symmetry represented locally by \(R\):
\begin{equation}
  \label{eq:74}
  \bigotimes_{v\in V}R_v(g) \circ \mathcal{G}_1 = \mathcal{G}_1.
\end{equation}

\subsection{Emergent (3+1)D quantum double models}

As established in Subsection \ref{sec:gauging-g-symmetries},
when gauging the global \(G\) symmetry of a state \(\ket{\psi}\)
in 2+1 dimensions,
the resulting state \(\mathcal{G}_0\ket{\psi}\)
exhibits an emergent 1-form \(\Rep G\) symmetry of the type considered above,
hence, we can apply the gauging procedure \(\mathcal{G}_1\),
which again results in a state with a global \(G\) symmetry.
Iterating this procedure yields the state
\begin{equation}
  \label{eq:78}
  \mathcal{G}\coloneq
  \cdots
  \circ \mathcal{G}_1\circ\mathcal{G}_0\circ \mathcal{G}_1\circ \mathcal{G}_0
  \ket{\psi},
\end{equation}
which can be viewed as having degrees of freedom on the edges of
a 3-dimensional cubic lattice.

As in the (1+1)-dimensional case, the state \(\mathcal{G}\ket{\psi}\) has \(G\) symmetries,
encoded by star operators and labeled by vertices,
and \(\Rep G\) symmetries,
encoded by plaquette operators and labeled by faces of the cubic lattice.
In the bulk, the star operators are given by
\begin{equation}
  \label{eq:77}
  S_v(g) \coloneq \input{figures/StarOperator3Dlattice},
\end{equation}
while the plaquette operators \(P_f\)
remain the same as in the 1+1 dimesional case:
\begin{align}
  \label{eq:76}
  \input{figures/PlaquetteOperator3dXY},
  &&\input{figures/PlaquetteOperator3dXZ},
  &&\input{figures/PlaquetteOperator3dYZ}.
\end{align}

These symmetries correspond to the stabilizer
terms of the (3+1)-dimensional quantum double models. Hence, the state \(\mathcal{G}\ket{\psi}\) is a ground state of the
bulk Hamiltonian
\begin{equation}
  \label{eq:H3D}
  H \coloneq -\sum_v A_v -\sum_f B_f,
\end{equation}
where
\begin{align}
  \label{eq:h3D}
  A_v \coloneq \frac{1}{|G|}\sum_{g\in G} S_v(g),&&
  B_f \coloneq \frac{1}{|\Rep G\,|}\sum_{\rho} P_f(\rho).
\end{align}

\section{Conclusion}
\label{sec:conclusion}

In this work, we have shown that every (2+1)-dimensional quantum double model based on groups can be reconstructed from its boundary through an iterative application of gauging maps.
Previous approaches were limited to abelian symmetry groups
because it was not known how to iterate the gauging procedure in the non-abelian case.
We overcome this limitation by identifying the correct class of Frobenius algebras internal to \(\Rep G\),
leading to full untwisted gauging of \(\Rep G\) symmetries
and determining explicitly the emergent dual group symmetry after gauging.
This clarifies, in concrete tensor-network terms, how gauging and duality intertwine in one dimension
and how their iteration naturally yields the commuting-projector Hamiltonians of Drinfeld doubles.

Further extending our formalism to two spatial dimensions,
we provided a gauging prescription for 1-form \(\Rep G\) symmetries,
returning a global \(G\) symmetry upon gauging,
and we showed that similarly iterating these procedures yields the (3+1)D quantum double models.
More generally, the gauging procedure for 1-form \(\Rep G\) symmetries proposed in this paper
is also valid for \((d-1)\)-form \(\Rep G\) symmetries in \(d\) spatial dimensions.

Our construction could provide a practical route for engineering and simulating non-abelian topological phases
from simple SPT or symmetry-broken boundary states \cite{lyonsProtocolsCreatingAnyons2025}.
Since such phases constitute key resources for topological quantum computation,
the explicit and constructive nature of our framework may inspire new approaches
to realizing non-abelian anyons in synthetic quantum platforms.
More broadly, our results support the view that categorical symmetries
form the natural language in which emergent gauge structures and topological orders should be understood.

A natural direction for future work is to extend these results to systems with 
\(H\) and \(\Rep H\) symmetries, where \(H\) is a semi-simple Hopf algebra,
since it is commonly believed that all non-anomalous global symmetries are Hopf algebra symmetries.
Another open question is the derivation of the bulk structure when twisted gauging of \(\Rep G\)
is used in the iterative gauging procedure,
for which an explicit description of the emergent dual symmetry is needed.
Significant progress on both problems is already being made.

A completely open question in this approach is
how to obtain quantum doubles with anomalous boundary symmetries,
such as the double-semion model.
Since conventional gauging procedures are not applicable in this setting,
resolving this issue may reveal an entirely new family of transformations.

\section{Acknowledgments}
D.B.\ thanks Andr\'as Moln\'ar for several insightful discussions and
Shuhei Ohyama for reminding us of reference~\cite{kapustinTopologicalFieldTheory2017}.
This research was funded in part by the European Union’s Horizon 2020
research and innovation programme (Grant No.\ \href{https://doi.org/10.3030/863476}{863476}),
and by the Austrian Science Fund (FWF) through the
Quantum Austria Funding Initiative project ``Entanglement Order Parameters''
(\href{https://doi.org/10.55776/P36305}{10.55776/P36305}),
as part of the European Union's NextGenerationEU instrument.
J.G.R.\ also acknowledges funding by the FWF Erwin Schrödinger Program (\href{https://doi.org/10.55776/J4796}{10.55776/J4796}).
\end{multicols}
\newpage

\begin{appendices}
\begin{multicols}{2}
\section{Computations}
\label{sec:computations}

In this section we give more details on the computation of Equations~\eqref{eq:48}.
From the definitions
\begin{align}
  
  \pgfkeys{
    /TNs/PEPS/.cd,
    width=2, 
    height=2, 
    label=\(\), 
    label inner=\(\),
    pattern = {},
    map top = 1,
    map LU = 1,
    map LD = 1,
    map RU = 1,
    map RD = 1,
    label = \(T_{\mathrm{o}}\), pattern=A
  }
  \input{figures/peps-tensor}
\,&\coloneq\sum_{g\in G}\,\input{figures/DefA},\\
  
  \pgfkeys{
    /TNs/PEPS/.cd,
    width=2, 
    height=2, 
    label=\(\), 
    label inner=\(\),
    pattern = {},
    map top = 1,
    map LU = 1,
    map LD = 1,
    map RU = 1,
    map RD = 1,
    label = \(\tw{\tau}{T}_{\mathrm{e}}\), pattern=B
  }
  \input{figures/peps-tensor}
\,&\coloneq\sum_{g\in G}\,\input{figures/DefB}
\end{align}
and
\begin{equation}
  \input{figures/RepGMPS}
  \coloneq \sum_{g\in G}
  \mStk[1.1em]{{}
               {
  \pgfkeys{
    /TNs/maps/.cd,
    label in=\(\), 
    label out=\(\), 
    width=1cm, 
    color=black,
    rev=0,
    arrows=mid arrow,
    map 1=1,
    map 2=1,
    map 3=1,
    pos 1=below,
    pos 2=below,
    pos 3=below,
    map 1 = \(\rho(g)\), pos 1 = above, color=violet
  }
  \input{figures/tn-map-hz}
}
               {
  \pgfkeys{
    /TNs/dyad/.cd,
    label=\(\), 
    width=.5cm, 
    height=1.3cm, 
    pivot=1,
    element=g,
    
  }
  
\begin{tikzpicture}[baseline={([yshift=-\the\fontdimen22\textfont2]D.center)}, rotate=-90, transform shape] 
\if1\RotatePoint
  \node[empty] (D) at (0, 0) {\scriptsize\(\Bra{\rotatebox{90}{\(\el\)}}\)};
\else
  \node[empty] (D) at (0, 0) {\scriptsize\(\Bra{\rotatebox[origin = \RotatePoint]{90}{\(\el\)}}\)};
\fi
\node[empty] (Hr) at (\heightTN/3, 0) {\TNLabel};
\draw[mid arrow] (Hr) -- ([xshift=-\the\fontdimen10\textfont3]D.east);
\node[empty] (lb) at (0, -\widthTN/2) {};
\node[empty] (rb) at (0,  \widthTN/2) {};
\end{tikzpicture}

}}
\end{equation}
it immediately follows that
\begin{subequations}\label{eq:63}
\begin{align}
  
  \pgfkeys{
    /TNs/PEPS/.cd,
    width=2, 
    height=2, 
    label=\(\), 
    label inner=\(\),
    pattern = {},
    map top = 1,
    map LU = 1,
    map LD = 1,
    map RU = 1,
    map RD = 1,
    label = \(\tw{\tau}{T}_{\mathrm{e}}\), pattern=B, map top = \(\tw{\overline{\tau}}{L}_g\)
  }
  \input{figures/peps-tensor}
&=
  
  \pgfkeys{
    /TNs/PEPS/.cd,
    width=2, 
    height=2, 
    label=\(\), 
    label inner=\(\),
    pattern = {},
    map top = 1,
    map LU = 1,
    map LD = 1,
    map RU = 1,
    map RD = 1,
    label = \(\tw{\tau}{T}_{\mathrm{e}}\), pattern=B, map LD = \(\tw{\tau}{L}_g^{\dag}\), map LU = \(L_g^{\dag}\)
  }
  \input{figures/peps-tensor}
,\\
  
  \pgfkeys{
    /TNs/PEPS/.cd,
    width=2, 
    height=2, 
    label=\(\), 
    label inner=\(\),
    pattern = {},
    map top = 1,
    map LU = 1,
    map LD = 1,
    map RU = 1,
    map RD = 1,
    label = \(\tw{\tau}{T}_{\mathrm{e}}\), pattern=B, map top = \(\tw{\tau}{R}_g\)
  }
  \input{figures/peps-tensor}
&=
  
  \pgfkeys{
    /TNs/PEPS/.cd,
    width=2, 
    height=2, 
    label=\(\), 
    label inner=\(\),
    pattern = {},
    map top = 1,
    map LU = 1,
    map LD = 1,
    map RU = 1,
    map RD = 1,
    label = \(\tw{\tau}{T}_{\mathrm{e}}\), pattern=B, map RD = \(\tw{\tau}{L}_g\), map RU = \(L_g\)
  }
  \input{figures/peps-tensor}
,\\
  
  \pgfkeys{
    /TNs/PEPS/.cd,
    width=2, 
    height=2, 
    label=\(\), 
    label inner=\(\),
    pattern = {},
    map top = 1,
    map LU = 1,
    map LD = 1,
    map RU = 1,
    map RD = 1,
    label = \(T_{\mathrm{o}}\), pattern=A, map top = \(L_g\)
  }
  \input{figures/peps-tensor}
&=
  
  \pgfkeys{
    /TNs/PEPS/.cd,
    width=2, 
    height=2, 
    label=\(\), 
    label inner=\(\),
    pattern = {},
    map top = 1,
    map LU = 1,
    map LD = 1,
    map RU = 1,
    map RD = 1,
    label = \(T_{\mathrm{o}}\), pattern=A, map LD = \(\tw{\alpha}{L}_g^{\dag}\), map RD = \(\tw{\alpha}{L}_g^{\vphantom{\dag}}\)
  }
  \input{figures/peps-tensor}
,\\
  
  \pgfkeys{
    /TNs/PEPS/.cd,
    width=2, 
    height=2, 
    label=\(\), 
    label inner=\(\),
    pattern = {},
    map top = 1,
    map LU = 1,
    map LD = 1,
    map RU = 1,
    map RD = 1,
    label = \(T_{\mathrm{o}}\), pattern=A, map top = \(R_g\)
  }
  \input{figures/peps-tensor}
&=
  
  \pgfkeys{
    /TNs/PEPS/.cd,
    width=2, 
    height=2, 
    label=\(\), 
    label inner=\(\),
    pattern = {},
    map top = 1,
    map LU = 1,
    map LD = 1,
    map RU = 1,
    map RD = 1,
    label = \(T_{\mathrm{o}}\), pattern=A, map LU = \(\tw{\alpha}{L}_g^{\dag}\), map RU = \(\tw{\alpha}{L}_g^{\vphantom{\dag}}\)
  }
  \input{figures/peps-tensor}
,
\end{align}
\end{subequations}
where \(\alpha\in \operatorname{H}^2(G, \operatorname{U}(1))\) is arbitrary
and
\begin{subequations}\label{eq:64}
\begin{align}
  \input{figures/trafoO1} &= \input{figures/trafoO2},\\
  \input{figures/trafoO3} &= \input{figures/trafoO4},\\
  \input{figures/trafoE1} &= \input{figures/trafoE2},\\
  \input{figures/trafoE3} &= \input{figures/trafoE4}.
\end{align}
\end{subequations}
\section{Internal Frobenius algebras}
\label{sec:frob-algebras}

\begin{definition}[Frobenius algebra]
  A \emph{Frobenius algebra} internal to a monoidal category~\(\mathcal{C}\)
  is an object~\(F\) in~\(\mathcal{C}\) together with morphisms
  \begin{align}
    \label{eq:59}
    \eqb{25.1}{\mu \colon F\otimes F\longrightarrow F,} &&
    \eqb{25.2}{\Delta \colon F\longrightarrow F\otimes F,}
  \end{align}
  termed multiplication and comultiplication, respectively, and
  \begin{align}
    \label{eq:69}
    \eqb{25.1}{\eta\colon \mathbbm{1}\longrightarrow F,}&&
    \eqb{25.2}{\varepsilon\colon F\longrightarrow \mathbbm{1},}
  \end{align}
  termed unit and counit, respectively, that define algebra 
  \begin{subequations}
  \begin{align}
    \label{eq:39}
    \mu\circ(\operatorname{id}\otimes\eta)
    = \eqb{39.1}{\operatorname{id}}
    &= \mu\circ(\eta\otimes\operatorname{id}),\\
    \mu \circ (\operatorname{id} \otimes\, \mu)
    &= \mu \circ (\mu\otimes\operatorname{id})
  \end{align}
  \end{subequations}
  and coalgebra structures 
  \begin{subequations}
  \begin{align}
    \label{eq:4}
    (\operatorname{id}\otimes\eta)\circ\Delta
    = \operatorname{id}
    &= (\eqb{4}{\eta\otimes\operatorname{id}})\circ\Delta,\\
    (\operatorname{id}\otimes\, \Delta)\circ \Delta
    &= (\eqb{4}{\Delta \otimes \operatorname{id}})\circ \Delta
  \end{align}
  \end{subequations}
  on \(F\)
  and satisfy the compatibility equation
  \begin{equation}
    \label{eq:2}
    (\mu\otimes\operatorname{id})\circ(\operatorname{id}\otimes\Delta)
    = \Delta\circ \mu
    = (\operatorname{id}\otimes\mu)\circ(\Delta\otimes\operatorname{id}).
  \end{equation}
\end{definition}

\begin{definition}[Symmetric Frobenius algebra]
  A Frobenius algebra \(F\) internal to a symmetric monoidal category~\(\mathcal{C}\)
  is called \emph{symmetric} if
  \begin{equation}
    \label{eq:49}
    \varepsilon \circ \mu \circ S_{F,F} = \varepsilon \circ \mu,
  \end{equation}
  where \(S\) is the symmetry in \(\mathcal{C}\).
\end{definition}
\begin{definition}[Haploid Frobenius algebra]
  A Frobenius algebra \(F\) internal to a \(\mathbb{C}\)-linear monoidal category
  is called \emph{haploid} \cite{fuchsCategoryTheoryConformal2006} if
  \begin{equation}
    \label{eq:44}
    \operatorname{Hom}(\mathbb{C}, F) \cong \mathbb{C}.
  \end{equation}
\end{definition}
\begin{definition}[Special Frobenius algebra]
  A Frobenius algebra internal to a \(\mathbb{C}\)-linear monoidal category
  is called \emph{special} if there exist nonzero \(\beta_F, \beta_{\mathbbm{1}} \in \mathbb{C}\) such that
  \begin{align}
    \label{eq:50}
    \mu \circ \Delta = \beta_F \operatorname{id},&&
    \epsilon\circ\eta = \beta_{\mathbbm{1}}.
  \end{align}
\end{definition}
\section{Graphs}\label{sec:graphs}
The following two definitions are copied from \cite{blanikInternalStructureGaugeinvariant2025}
and follow conventions established in \cite{serreTrees1980}.
\begin{definition}[Directed graph]
  A \emph{directed graph} is an ordered pair \(\Gamma = (V, E)\) of sets \(V\) and \(E\subseteq V\times V\).
  The elements of the set \(\operatorname{vert}\Gamma := V\) are called \emph{vertices} and
  the elements of the set \(\operatorname{edge}\Gamma := E\),
  which are ordered pairs of vertices, are called \emph{edges}.
  For any edge \(e\equiv (v_1,v_2) \in E\),
  the vertex \(\operatorname{o}(e) := v_1 \in V\) is called the \emph{origin} of \(e\) and
  the vertex \(\operatorname{t}(e) := v_2 \in V\) is called the \emph{terminus} of \(e\).
  For any vertex \(v\in V\) let
  \(E_v^{\mathrm{o}}\) denote the set of edges
  that have \(v\) as their origin
  (i.e.\ the outgoing edges),
  \(E_v^{\mathrm{t}}\) denote the set of edges
  that have \(v\) as their terminus
  (i.e.\ the incoming edges)
  and \(E_v\) denote the set of all adjacent edges.
  In formulae these sets are given as
  \begin{subequations}
  \begin{align}
  E_v^{\mathrm{o}} &:= \left\{ e\in E \mid v = \operatorname{o}(e)\right\},\\
  E_v^{\mathrm{t}} &:= \left\{ e\in E \mid v = \operatorname{t}(e)\right\},\\
  E_v &:= E_v^{\mathrm{o}} \cup E_v^{\mathrm{t}}.
  \end{align}
  \end{subequations}
\end{definition}
\begin{definition}[Oriented graph]
  A directed graph \(\Gamma\) for which
  \((v_1, v_2) \in \operatorname{edge}\Gamma \) implies
  \((v_2, v_1) \notin \operatorname{edge}\Gamma\),
  is called an \emph{oriented graph}.
\end{definition}
\begin{definition}[Path]
  Let \(\Gamma = (V, E)\) be a directed graph.
  A (finite) \emph{path}\footnote{
    What we call path is usually referred to as \emph{walk} in the literature.
  } in \(\Gamma\) is a sequence of vertices
  \((v_1, \ldots, v_n)\in V^n\)
  such that for all \(i = 1, \ldots, n-1\),
  either \((v_i, v_{i+1})\in E\) or \((v_{i+1}, v_i)\in E\).
  A path is called \emph{closed} if \(v_n = v_1\).
\end{definition}
\end{multicols}
\end{appendices}

\printbibliography
\end{document}